\documentclass{article}
\usepackage{spconf,amsmath,graphicx}

\usepackage{kotex} 
\usepackage{soul, color} 

\title{CAU\_KU team's submission to ADD 2022 Challenge task 1: Low-quality fake audio detection through frequency feature masking}
\twoauthors
  {Il-Youp Kwak\sthanks{This work was supported by the National Research Foundation of Korea (NRF) grant funded by the Korea government(MSIP) (No. NRF-2020R1C1C1A01013020).}, Jonghoon Yang, Yerin Lee}
	{Chung-Ang University\\
	Seoul, Korea}
  {Sunmook Choi, Seungsang Oh\sthanks{This work was supported by the National Research Foundation of Korea(NRF) grant funded by the Korea government(MSIP) (No. NRF-2017R1A2B2007216).}}
	{Korea University\\%
	Seoul, Korea}
\begin{document}
%
\maketitle
\begin{abstract}
This technical report describes Chung-Ang University and Korea University (CAU\_KU) team's model participating in the Audio Deep Synthesis Detection (ADD) 2022 Challenge, track 1: Low-quality fake audio detection. For track 1, we propose a frequency feature masking (FFM) augmentation technique to deal with a low-quality audio environment. 
We applied FFM and mixup augmentation on five spectrogram-based deep neural network architectures that performed well for spoofing detection using mel-spectrogram and constant Q transform (CQT) features. Our best submission achieved 23.8\% of EER ranked 3rd on track 1.
\end{abstract}

\begin{keywords}
Audio deep synthesis, low-quality audio, deep learning, frequency feature masking, audio data augmentation
\end{keywords}

\section{Introduction}
\label{sec:intro}

This technical paper describes Chung-Ang University and Korea University (CAU\_KU) team's participation in the Audio Deep Synthesis Detection (ADD) 2022 Challenge, track 1 on Low-quality fake audio detection. The dataset for track1 comprises bonafide and fake utterances. The utterance samples are generated using text-to-speech and voice conversion algorithms with various real-world noises such as background music effects. The goal of track 1 is to derive a model that classifies bonafide and fake utterances. For the system development, organizers provided adaptation set of bonafide and fake utterances in a noisy environment. 

For track 1, we propose a frequency feature masking (FFM) augmentation for robust training in a noisy environment.
FFM is similar to the SpecAugment \cite{48482} in that the augmentation policy comprises wraping the features, masking blocks of frequency channels or time steps. FFM is designed to focus more on voice spoofing problems in a noisy environment. We applied FFM and mixup \cite{zhang2018mixup} augmentation on five spectrogram-based deep neural network architectures that performed well for spoofing detection using mel-spectrogram and constant Q transform (CQT) features. Five systems are 1) ResMax \cite{kwak2021} with CQT feature, 2) Light CNN (LCNN) \cite{lcnn, asvspoof2019p2} with CQT feature, 3) BC-ResMax (a variant of BC-ResNet \cite{kim2021broadcasted}) with mel-spectrogram feature, 4) Double Depthwise Separable net(DDWSnet) with mel-spectrogram feature, and 5) Overlapped Frequency-distributed (OFD) model with CQT feature. We slightly modified the existing models \cite{kwak2021, asvspoof2019p2} for systems 1) and 2). The 3), 4), and 5) systems are our manually built versions and are defined in the methods section. We ensembled those five systems for the final submission. 












\section{Methods}

\subsection{Feature engineering}
We utilized CQT and mel-spectrogram feature extracted using the librosa software \cite{mcfee2015librosa}. For the CQT feature extraction, we set the minimum frequency as 5, the number of frequency bins as 100, and the filter scale factor as 1. For the mel-spectrogram feature extraction, we used two different options. The first one is 100 frequency bins with 1024 window lengths and 512 hop sizes. The second one is 120 frequency bins with 2048 window lengths and 1024 hop sizes. 

\subsection{Mixup augmentation}

Mixup is a widely used data augmentation in voice classifications as well as image classifications \cite{zhang2018mixup}. We mixed different samples of the training set according to a parameter $\lambda$ which is sampled from the $beta(\alpha, \alpha)$ distribution with a parameter $\alpha$. In our modeling, we set $\alpha$ from 0.4-0.9. The method is as follows:

\begin{equation}
    X = \lambda X_i + (1-\lambda)X_j,
\end{equation}
\begin{equation}
    y = \lambda y_i + (1-\lambda)y_j,
\end{equation}

\noindent where $X_i$ and $X_j$ are different spectrogram images with their corresponding labels, $y_i$ and $y_j$, respectively. 

\subsection{Frequency feature masking (FFM) augmentation}
\begin{figure}[htb]

\begin{minipage}[b]{.48\linewidth}
  \centering
  \centerline{\includegraphics[width=4.0cm]{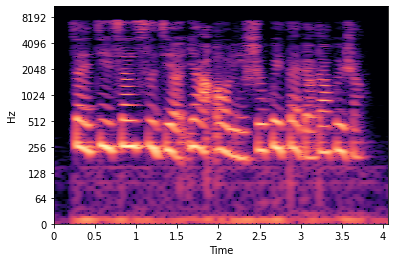}}
  \centerline{(a) Normal environment}\medskip
\end{minipage}
\hfill
\begin{minipage}[b]{0.48\linewidth}
  \centering
  \centerline{\includegraphics[width=4.0cm]{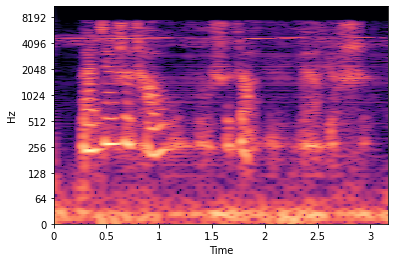}}
  \centerline{(b) Noisy environment}\medskip
\end{minipage}
\caption{A genuine voice sample in a normal environment from training data (a), and in a noisy environment from adaptation data (b).}
\label{fig:spec1}
\end{figure}

Figure \ref{fig:spec1}(a) and (b) show a genuine voice sample in a normal environment from training data and in a noisy environment from adaptation data, respectively. 
From Figure \ref{fig:spec1}(b), we can find long horizontal lines in 4096 Hz regions. The utterance sample had a fixed high-frequency noise signal when we heard the utterance sample.
We can think of the following scenarios:
\begin{enumerate}
    \item High-frequency areas may have more noise. Let the model focus more on other frequency areas by masking high-frequency areas.
    \item Low-frequency areas may have more noise. Let the model focus more on other frequency areas by masking low-frequency areas.
    \item If the model is trained to focus on specific frequency bands, performance will decrease when there is a noise signal in the specific frequency band. We can utilize random frequency band masking for the balanced model training.
\end{enumerate}

\begin{figure}[htb]

\begin{minipage}[b]{.48\linewidth}
  \centering
  \centerline{\includegraphics[width=4cm]{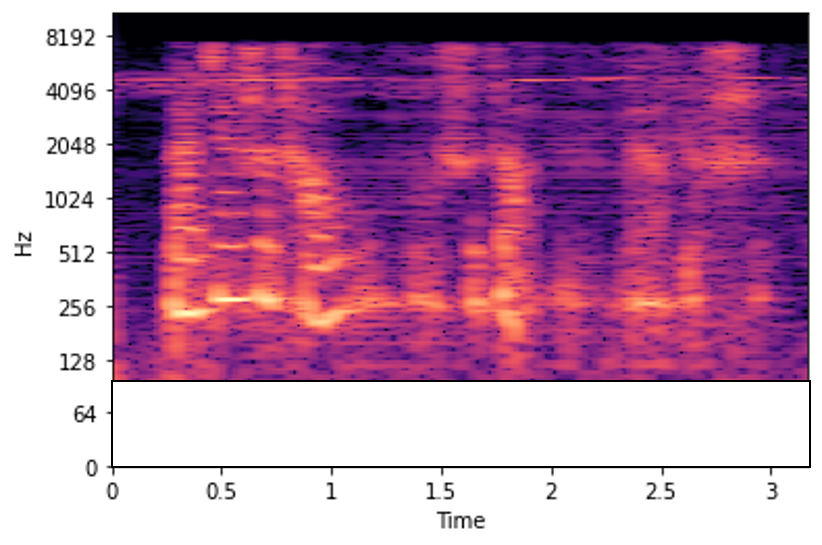}}
  \centerline{(a) Low-frequency masking}\medskip
\end{minipage}
\hfill
\begin{minipage}[b]{0.48\linewidth}
  \centering
  \centerline{\includegraphics[width=4cm]{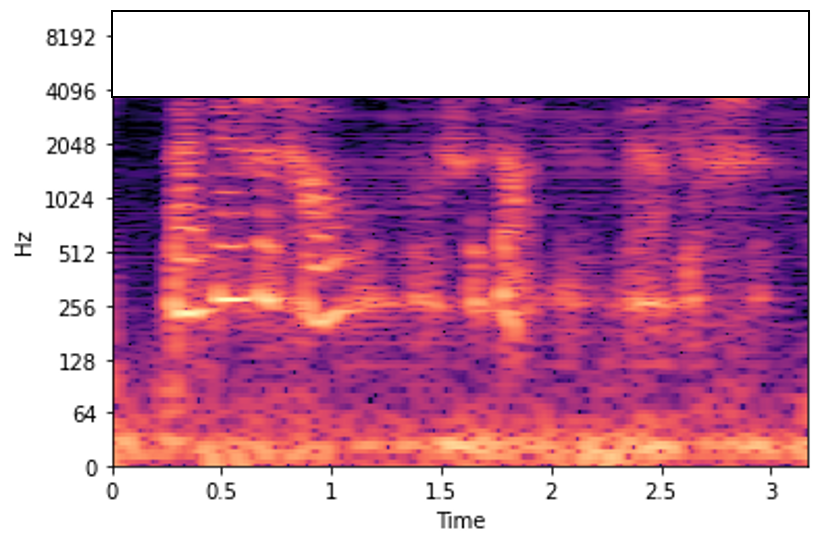}}
  \centerline{(b) High-frequency masking}\medskip
\end{minipage}
\hfill
\begin{minipage}[b]{1.0\linewidth}
  \centering
  \centerline{\includegraphics[width=4cm]{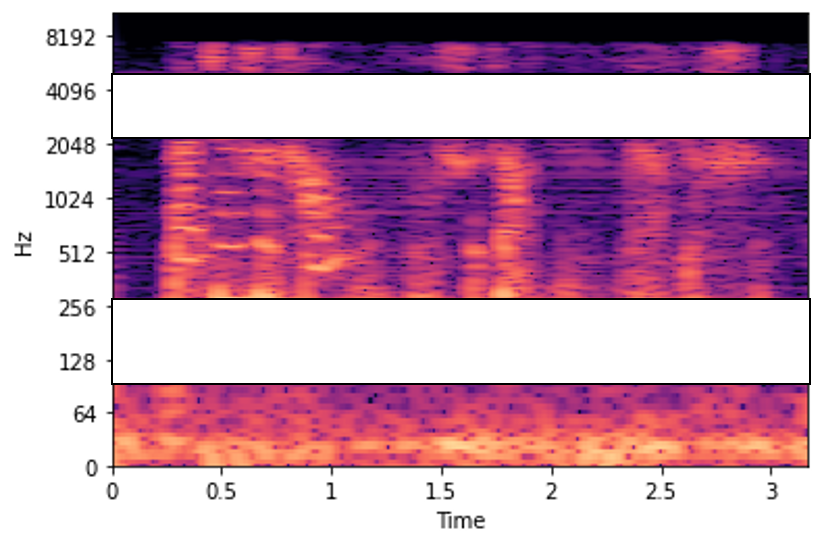}}
  \centerline{(c) Random frequency band masking}\medskip
\end{minipage}

\caption{Description on Low-frequency masking, High-frequency masking and Random frequency band masking.}
\label{fig:masking}
\end{figure}


%
%

To address the three scenarios, we considered three types of masking, `Low-frequency masking', `High-frequency masking', and `Random frequency band masking' (shown in Figure \ref{fig:masking}). 

In every training epochs, for a given training spectrogram sample, Low-frequency masking is applied with probability $p_l$ as described in Figure \ref{fig:masking}(a). The Low-frequency area is randomly selected (ex. from 0 to 14 range in 100 mel bins), and all values in the corresponding spectrogram values are set to 0. 

High-frequency masking described in Figure \ref{fig:masking}(b) is similar to Low-frequency masking. Only differences are the masking probability $p_h$ and selected frequency area (ex. from 85 to 100 range in 100 mel bins). 

Random frequency band masking is described in Figure \ref{fig:masking}(c). In this case, Random frequency band masking is applied with probability $p_r$, and the number of frequency bands to be masked out, band sizes, and locations are also randomly selected. Then, the values in the selected frequency bands are set to 0. In Figure \ref{fig:masking}(c), two frequency bands are masked out.










\subsection{Models}
In order to use the previously proposed FFM augmentation, five spectrogram-based models were considered. 

\subsubsection{LCNN model}
The LCNN model has proven useful in ASVspoof 2017, 2019, and 2021 competitions \cite{asvspoof2017p1, asvspoof2019p2, tomilov21_asvspoof,asvspoof2017, asvspoof2019, yamagishi21_asvspoof}. 
We used a deeper LCNN model by adding a few more layers to the Light CNN-9 model \cite{lcnn}. Light CNN-9 model repeats five convolution layers and four network-in-network (NIN) layers \cite{lcnn}. We proposed a model which iterates six convolution layers and 5 NIN layers using 32, 48, 64, 32, 32, and 32 convolution filters and 32, 48, 64, 64, and 32 NIN filters. As in the previous model, the kernel size of the first convolution layer is set to 5, and the remaining convolution layers are set to 3. Except for the first and third convolution layers, batch normalizations are followed. All NIN layers are followed by batch normalization layer. 
Instead of using a fully connected layer defined in the Light CNN-9 model \cite{lcnn}, we used the global average pooling layer, batch normalization, and Dropout layer with a probability of 0.5.


\subsubsection{ResMax model}
We previously proposed the ResMax model and confirmed that it showed excelent performance in the ASVspoof 2019 competition data \cite{kwak2021}. We used the same model.

\subsubsection{Double Depthwise Separable (DDWS) model}
\begin{figure}[htb]
  \centering
  \centerline{\includegraphics[width=9cm]{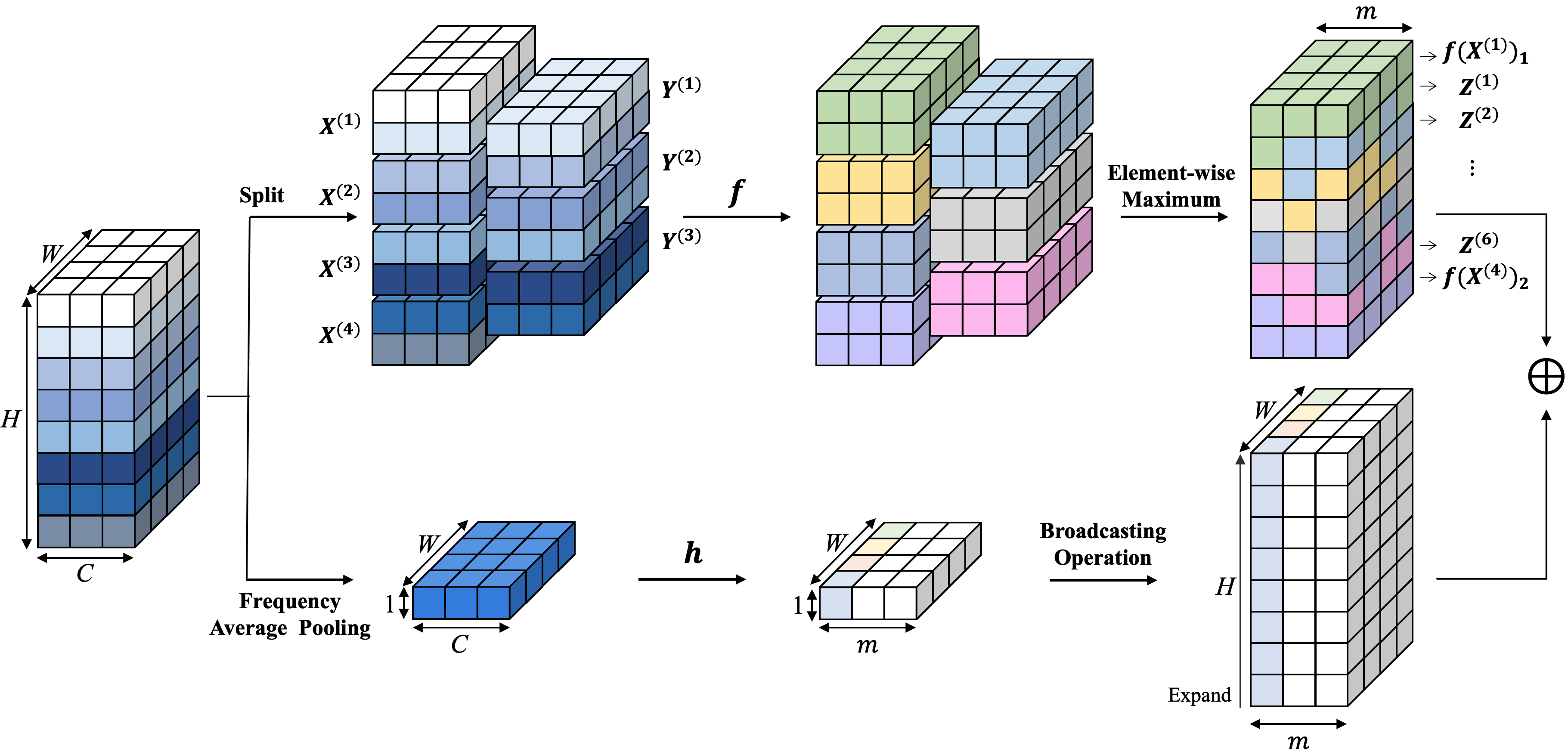}}
  \centerline{(c) OFD block when $n=4$}\medskip
\caption{Model Blocks}
\label{fig:blocks}
\end{figure}

\begin{figure}[htb]
\begin{minipage}[b]{.48\linewidth}
  \centering
  \centerline{\includegraphics[width=3.7cm]{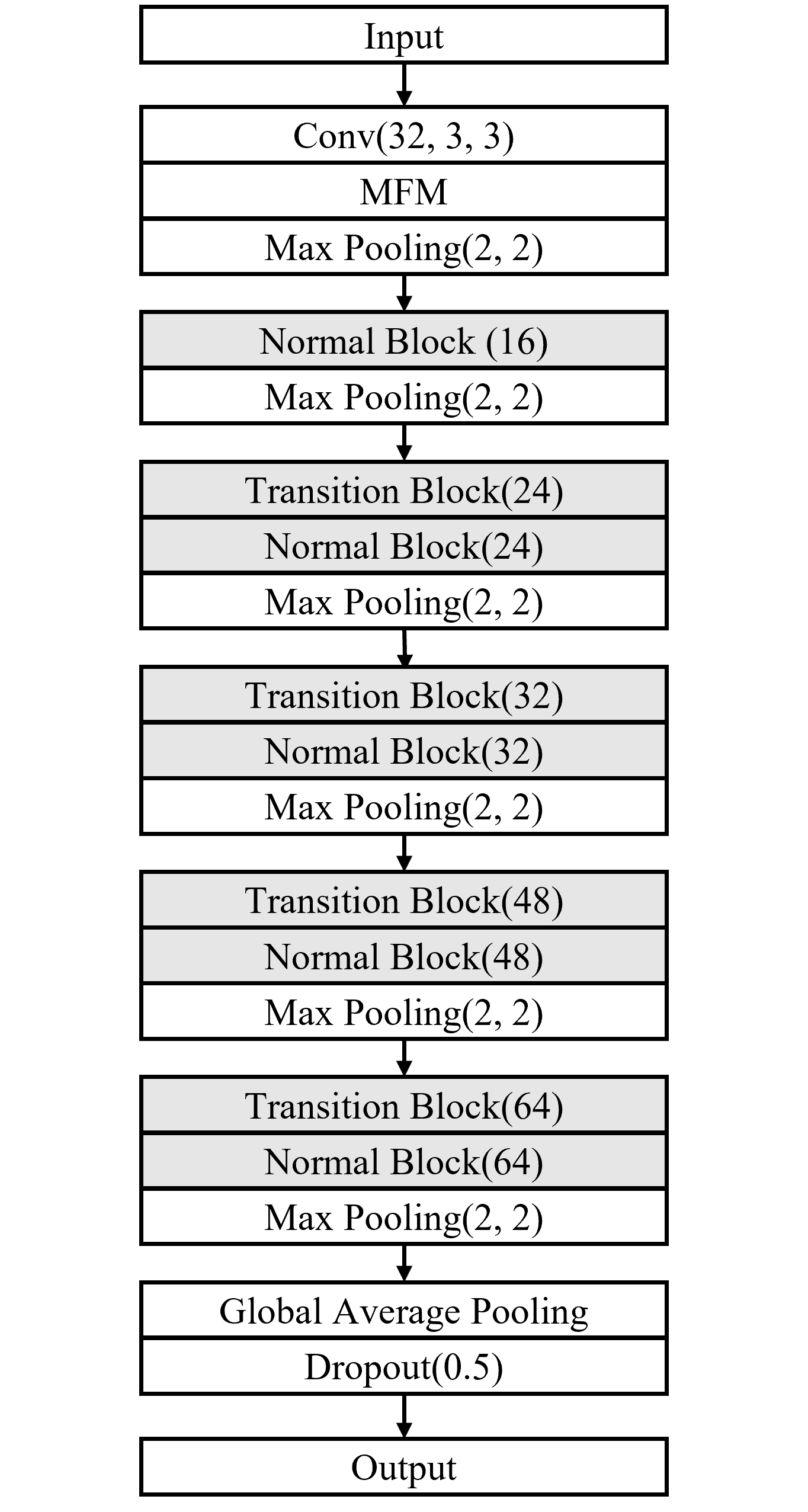}}
  \centerline{(a) DDWS/BC-ResMax model}\medskip
\end{minipage}
\begin{minipage}[b]{0.48\linewidth}
  \centering
  \centerline{\includegraphics[width=3.0cm]{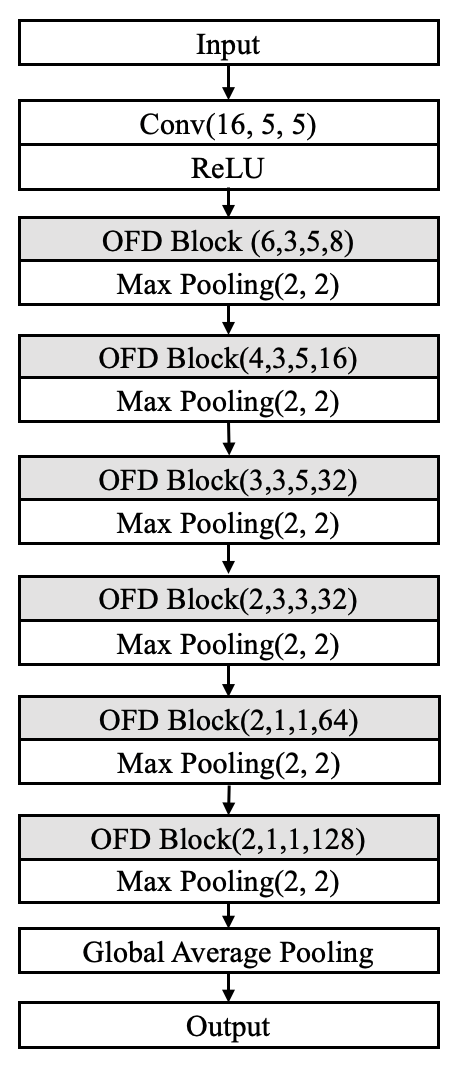}}
  \centerline{(b) OFD model}\medskip
\end{minipage}
\caption{Model Architectures}
\label{fig:model archi}
\end{figure}

The $(k_1, k_2)$ filter is used as the depthwise convolution filter in the original depthwise separable convolution \cite{howard2017mobilenets, 8099678}. We propose to apply the existing depthwise separable convolution as two depthwise convolutions with filter size $(k_1,1)$ and $(1,k_2)$ to consider frequency information and temporal information separately.
Precisely, we define DDWS block as shown in Figure \ref{fig:blocks}(a). We define $f_{1}$ to be depthwise convolutions with $(1, k_2)$ filter followed by Subspectral Normalization (SSN) \cite{chang2021subspectral} and Swish activation \cite{Ramachandran2017SwishAS}. We define $f_{2}$ to be depthwise convolutions with $(k_1, 1)$ filter followed by SSN and ReLU activation. 
Lastly, define $g$ as a composite of pointwise convolution, ReLU activation, and spatial dropout.
A block design without dotted marks on the top is a normal block, and a design with dotted marks is a transition block. The full model architecture is described in Figure \ref{fig:model archi}(a).


\begin{figure}[htb]
\begin{minipage}[b]{.48\linewidth}
  \centering
  \centerline{\includegraphics[width=3.5cm]{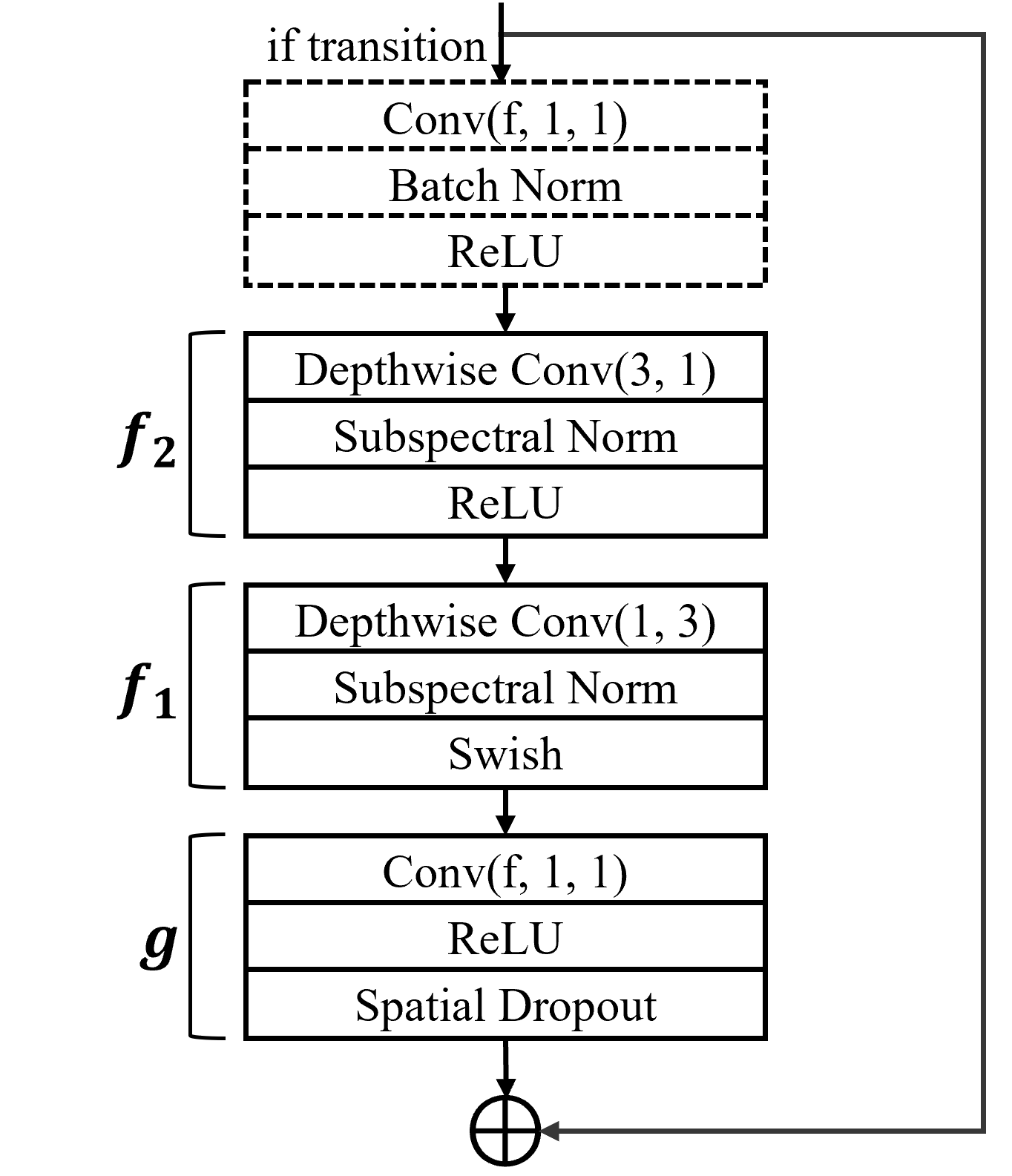}}
  \centerline{(a) DDWS block}\medskip
\end{minipage}
\begin{minipage}[b]{0.48\linewidth}
  \centering
  \centerline{\includegraphics[width=3.3cm]{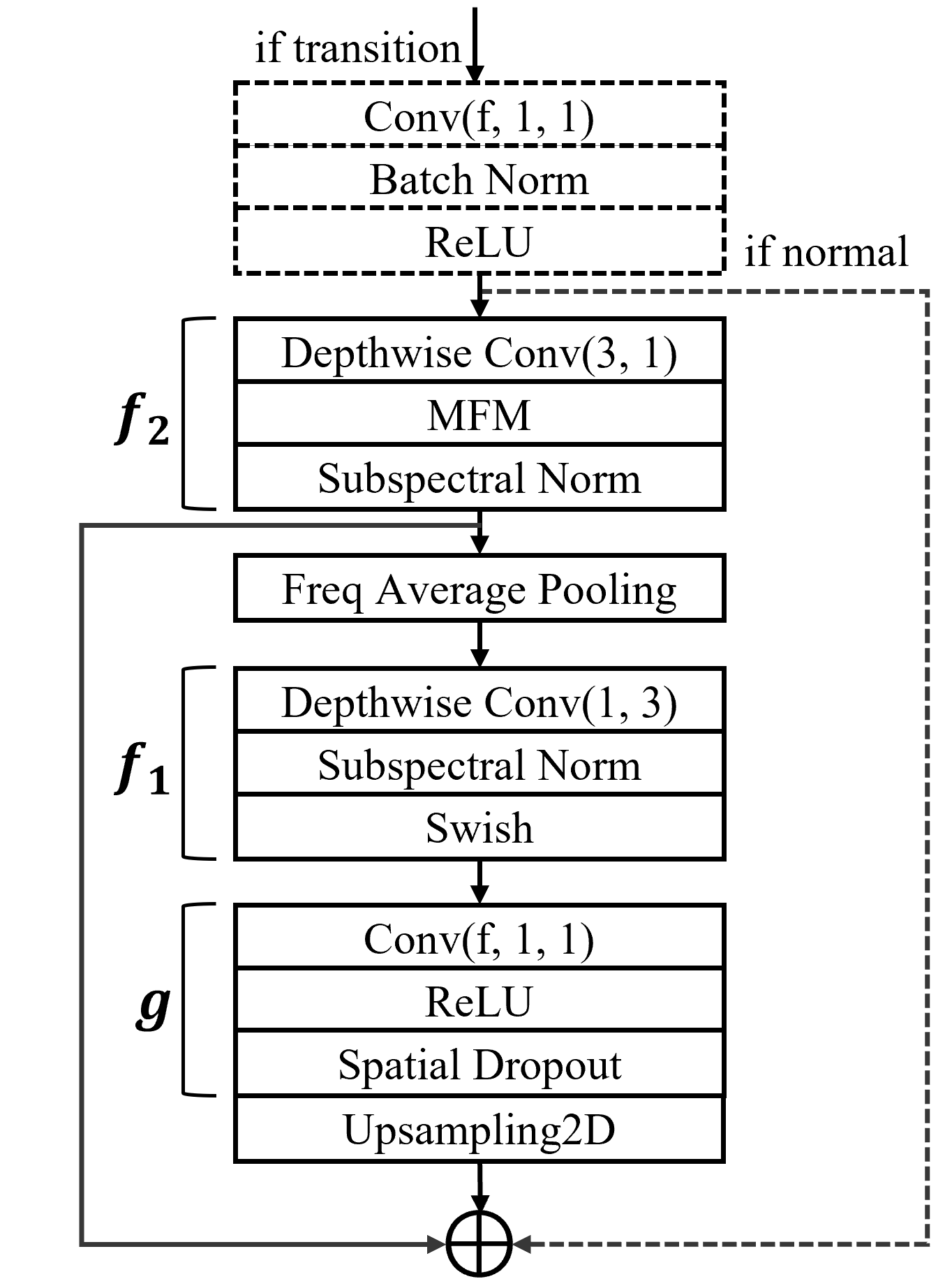}}
  \centerline{(b) BC-ResMax block}\medskip
\end{minipage}
\end{figure}
%

\subsubsection{BC-ResMax model}
We revised BC-ResNet \cite{Kim2021b, kim2021broadcasted} by integrating max feature map (MFM) activation from LCNN. Our proposed BC-ResMax block is described in Figure \ref{fig:blocks}(b). We define $f_{2}$ to be depthwise convolutions with $(k_1, 1)$ filter followed by MFM and SSN. A block design without dotted marks on the top is a normal block, and a design with dotted marks is a transition block. The full model architecture is described in Figure \ref{fig:model archi}(a).




\subsubsection{Overlapped Frequency-Distributed (OFD) model}
We introduce a model called the `Overlapped Frequency-Distributed (OFD) model.' 
The model consists of 6 OFD blocks, each of which has two streams, described in Figure \ref{fig:blocks}(c). The first stream is inspired by the FreqCNN model \cite{Mao2019AudioCU} and the second stream is inspired by BC-ResNet model \cite{Kim2021b, kim2021broadcasted}.

The first stream is to learn frequency-related features. 
Let $X = (x_{ij})\in \mathbf{R}^{H\times W} $ be an input of the block, where $H$ and $W$ correspond to the frequency and temporal dimensions, respectively, and the channel dimension is omitted. 
Let $n$ be the number split along the frequency axis. If necessary, $X$ is zero-padded along the frequency axis so that $H$ is divisible by $2n$. Then, 
we split $X$ into $n$ disjoint parts $\{X^{(k)}\}_{k=1}^n$ and $n-1$ overlapped parts $\{Y^{(k)}\}_{k=1}^{n-1}$, which are as follows:
$$X^{(k)} = (x_{ij}) \in \mathbf{R}^{2s \times W}, \quad 1+2(k-1)s \leq i \leq 2ks$$
$$Y^{(k)} = (x_{ij}) \in \mathbf{R}^{2s \times W}, \quad 1+(2k-1)s \leq i \leq (2k+1)s$$
where $s = \frac{H}{2n}$. Note that the lower half part of $X^{(k)}$ and the upper half part of $Y^{(k)}$ coincide, and the upper half part of $X^{(k+1)}$ and lower half part of $Y^{(k)}$ coincide, which is why we call $Y^{(k)}$'s overlapped parts.
Define a function $f$ which is a composite of $k_1 \times 1$ convolution, batch normalization, ReLU, $k_1 \times 1$ convolution, and batch normalization.
Note that no activation is used in the second convolution, and zero-padding is required in each convolution to keep the size the same.
Next, $f(X^{(k)})$'s and $f(Y^{(k)})$'s should be joined to reconstruct the full 2D image of the size $H \times W$. Each of them is divided into 2 parts, which are upper and lower half parts, resulting $f(X^{(k)})_1$, $f(X^{(k)})_2$, $f(Y^{(k)})_1$, and $f(Y^{(k)})_2$.
Then, we define 
$Z^{(2k-1)} = \max \left\{f(X^{(k)})_2, f(Y^{(k)})_1\right\}$ and $Z^{(2k)} = \max \left\{f(X^{(k+1)})_1, f(Y^{(k)})_2\right\}$
where $\max$ is an element-wise maximum operation. Note that the maximum operation acts like an MFM activation, which justifies no activation in the last convolution in the function $f$.
Finally, we have a set of sub-images, 
$$\left\{f(X^{(1)})_1, Z^{(1)}, \ldots, Z^{(2n-2)}, f(X^{(n)})_2\right\},$$ 
each of which is of the size $s \times W$. Then, the final output of the stream is obtained by concatenating these $2n$ sub-images along the frequency axis.

The second stream is to learn temporal features. First, the input $X$ is averaged out along the frequency axis so that we get a feature map of size $\mathbf{R}^{1\times W}$. Next, we define a function $h$ which is a composite of $1 \times k_2$ depthwise convolution with dilation of 4, batch normalization, swish activation, $1 \times 1$ pointwise convolution with ReLU activation, and spatial dropout. Note that we use zero padding to keep the temporal dimension the same. Lastly, we expand the feature map along the frequency axis using broadcasting operations.

The same number $m$ of filters are used in convolutions for both streams, and the dropout rate is fixed to 0.5. Each OFD block is represented as OFD Block($n,k_1,k_2,m$) in the figure. The overall architecture of the OFD model is illustrated in Figure \ref{fig:model archi}(b).
In the model, we have set $k_1=k_2=1$ in the 5th and 6th blocks to control the receptive field in the network \cite{9439825}.


\section{Experiments}

\subsection{Experiments on data augmentation}
We experimented with our proposed models and augmentation techniques on the adaptation set which includes real-world noises and background music. Table \ref{exp_tb} describes EER on the adaptation set for each model with different augmentation methods applied. LF is Low-frequency masking, HF is High-frequency masking, and RF is Random frequency band masking. There is a larger difference in the applications of the augmentation techniques rather than the differences among the models. Without any data augmentations, BC-ResMax, DDWS, ResMax, and OFD models have EER of 22.53\%, 23.29\%, 21.40\%, and 22.66\%. Mixup augmentation decreased those EERs to 18.85\%, 19.62\%, 16.91\%, and 18.29\%. Additional FFM augmentation remarkably decreased EER of BC-ResMax, DDWS, and ResMax models to 11.89\%, 11.99\%, and 14.02\%. These results show that the proposed FFM augmentation is effective considering the noisy environment. OFD model didn't use FFM augmentation technique since the model separately finds features from different frequency ranges.

\begin{table}[ht]
\centering
\caption{EER on the adaptation set for each model with different augmentation methods. } \label{exp_tb}
\scalebox{0.93}{\begin{tabular}[t]{|l|c|c|c|c|c|r|}
\hline
 Model    &  Feature & Mixup & LF & HF & RF & EER \\
\hline
BC-ResMax  & CQT & X & X & X & X & 22.53\% \\ 
BC-ResMax  & CQT & O & X & X & X & 18.85\% \\ 
BC-ResMax  & CQT & O & O & X & X & 15.51\% \\ 
BC-ResMax  & CQT & O & X & O & X & 14.07\% \\ 
BC-ResMax  & CQT & O & O & O & O & 11.89\% \\ 
\hline
DDWS  & melspec & X & X & X & X & 23.29\% \\ 
DDWS  & melspec & O & X & X & X & 19.62\% \\ 
DDWS  & melspec & O & X & O & O & 11.99\% \\ 
DDWS  & melspec & O & O & X & O & 14.42\% \\ 
DDWS  & melspec & O & O & O & O & 12.62\% \\ 
\hline
ResMax  & CQT & X & X & X & X & 21.40\% \\ 
ResMax  & CQT & O & X & X & X & 16.91\% \\ 
ResMax  & CQT & O & O & X & O & 15.73\% \\ 
ResMax  & CQT & O & X & O & O & 15.49\% \\ 
ResMax  & CQT & O & O & O & O & 14.02\% \\ 
\hline
OFD  & CQT & X & X & X & X & 22.66\% \\ 
OFD  & CQT & O & X & X & X & 18.29\% \\ 
\hline

\end{tabular}}
\end{table}

\subsection{Submitted ensemble system}
Table \ref{res} describes five top-performing single systems, data augmentation methods applied, their EER on final evaluation data, weights for the final ensemble model, and the EER of our ensemble system. Applying all LF, HF, and RF augmentation in the final evaluation data does not always produce the best result. Thus, we have tested various combinations. The final models were selected to have as few correlations as possible, and ensemble weights were calculated based on EER. 

\begin{table}[h]
\centering
\caption{EER (\%) on the final evaluation data from the ADD challenge, and weights for ensemble model. } \label{res}
\scalebox{.9}{\begin{tabular}[t]{|l|l|c|c|c|}
\hline
 Model    &  Feature & Augmentation & EER & weights \\
\hline
LCNN & CQT & Mixup, LF, RF & 26.05\% & 0.20\\ 
ResMax  & CQT & Mixup, RF & 24.7\% & 0.27 \\ 
DDWS  & melspec & Mixup, RF & 26.40\%  & 0.20 \\ 
BC-ResMax  & melspec & Mixup, LF, RF & 27.34\% & 0.13 \\ 
OFD  & CQT & Mixup & 26.02\% & 0.20 \\ 
\hline
Ensemble  & - & - & 23.8\%  & -  \\ 
\hline
\end{tabular}}
\end{table}

\section{Conclusion}
This paper proposed FFM augmentation and three new models, DDWS, BC-ResMax, and OFD. Especially, FFM is beneficial in overcoming noisy environments. DDWS, BC-ResMax, and OFD showed similar performance to the existing spoofing detection models, LCNN and ResMax. Our final ensemble model comprising LCNN, ResMax, DDWS, BC-ResMax, and OFD achieved 23.8\% of EER, placing 3rd in the ADD competition, track 1.


%
%
%


\vfill\pagebreak

\bibliographystyle{IEEEbib}
\bibliography{main_track1}

\begin{thebibliography}{10}

\bibitem{48482}
Daniel~S. Park, William Chan, Yu~Zhang, Chung-Cheng Chiu, Barret Zoph,
  Ekin~Dogus Cubuk, and Quoc~V. Le,
\newblock ``Specaugment: A simple augmentation method for automatic speech
  recognition,''
\newblock in {\em Proc. Interspeech 2019}, 2019.

\bibitem{zhang2018mixup}
Hongyi Zhang, Moustapha Cisse, Yann~N. Dauphin, and David Lopez-Paz,
\newblock ``mixup: Beyond empirical risk minimization,''
\newblock in {\em International Conference on Learning Representations}, 2018.

\bibitem{kwak2021}
Il-Youp Kwak, Sungsu Kwag, Junhee Lee, Jun~Ho Huh, Choong-Hoon Lee, Youngbae
  Jeon, Jeonghwan Hwang, and Ji~Won Yoon,
\newblock ``{ResMax: Detecting Voice Spoofing Attacks with Residual Network and
  Max Feature Map},''
\newblock in {\em 25th International Conference on Pattern Recognition (ICPR)}.
  2021, pp. 4837--4844, {IEEE} Computer Society.

\bibitem{lcnn}
X.~{Wu}, R.~{He}, Z.~{Sun}, and T.~{Tan},
\newblock ``A light cnn for deep face representation with noisy labels,''
\newblock {\em IEEE Transactions on Information Forensics and Security}, vol.
  13, no. 11, pp. 2884--2896, Nov 2018.

\bibitem{asvspoof2019p2}
Galina Lavrentyeva, Sergey Novoselov, Andzhukaev Tseren, Marina Volkova, Artem
  Gorlanov, and Alexandr Kozlov,
\newblock ``{STC Antispoofing Systems for the ASVspoof2019 Challenge},''
\newblock in {\em Proc. Interspeech 2019}, 2019, pp. 1033--1037.

\bibitem{kim2021broadcasted}
Byeonggeun Kim, Simyung Chang, Jinkyu Lee, and Dooyong Sung,
\newblock ``Broadcasted residual learning for efficient keyword spotting,''
\newblock in {\em Proc. Interspeech 2021}, 2021, pp. 4538--4542.

\bibitem{mcfee2015librosa}
Brian McFee, Colin Raffel, Dawen Liang, Daniel~PW Ellis, Matt McVicar, Eric
  Battenberg, and Oriol Nieto,
\newblock ``librosa: Audio and music signal analysis in python,''
\newblock in {\em Proceedings of the 14th python in science conference}, 2015,
  vol.~8.

\bibitem{asvspoof2017p1}
Galina Lavrentyeva, Sergey Novoselov, Egor Malykh, Alexander Kozlov, Oleg
  Kudashev, and Vadim Shchemelinin,
\newblock ``Audio replay attack detection with deep learning frameworks,''
\newblock in {\em Proc. Interspeech 2017}, 2017.

\bibitem{tomilov21_asvspoof}
Anton Tomilov, Aleksei Svishchev, Marina Volkova, Artem Chirkovskiy, Alexander
  Kondratev, and Galina Lavrentyeva,
\newblock ``{STC Antispoofing Systems for the ASVspoof2021 Challenge},''
\newblock in {\em Proc. 2021 Edition of the Automatic Speaker Verification and
  Spoofing Countermeasures Challenge}, 2021, pp. 61--67.

\bibitem{asvspoof2017}
Tomi Kinnunen, Md~Sahidullah, Hector Delgado, Massimiliano Todisco, Nicholas
  Evans, Junichi Yamagishi, and Kong~Aik Lee,
\newblock ``The asvspoof 2017 challenge: Assessing the limits of replay
  spoofing attack detection,''
\newblock in {\em Proc. Interspeech 2017}, Stockholm, 2017, pp. 2--6, ISCA.

\bibitem{asvspoof2019}
Massimiliano Todisco, Xin Wang, Ville Vestman, Md. Sahidullah, Héctor Delgado,
  Andreas Nautsch, Junichi Yamagishi, Nicholas Evans, Tomi~H. Kinnunen, and
  Kong~Aik Lee,
\newblock ``{ASVspoof 2019: Future Horizons in Spoofed and Fake Audio
  Detection},''
\newblock in {\em Proc. Interspeech 2019}, 2019, pp. 1008--1012.

\bibitem{yamagishi21_asvspoof}
Junichi Yamagishi, Xin Wang, Massimiliano Todisco, Md~Sahidullah, Jose Patino,
  Andreas Nautsch, Xuechen Liu, Kong~Aik Lee, Tomi Kinnunen, Nicholas Evans,
  and Héctor Delgado,
\newblock ``{ASVspoof 2021: accelerating progress in spoofed and deepfake
  speech detection},''
\newblock in {\em Proc. 2021 Edition of the Automatic Speaker Verification and
  Spoofing Countermeasures Challenge}, 2021, pp. 47--54.

\bibitem{howard2017mobilenets}
Andrew~G. Howard, Menglong Zhu, Bo~Chen, Dmitry Kalenichenko, Weijun Wang,
  Tobias Weyand, Marco Andreetto, and Hartwig Adam,
\newblock ``Mobilenets: Efficient convolutional neural networks for mobile
  vision applications,''
\newblock {\em CoRR}, vol. abs/1704.04861, 2017.

\bibitem{8099678}
François Chollet,
\newblock ``Xception: Deep learning with depthwise separable convolutions,''
\newblock in {\em 2017 IEEE Conference on Computer Vision and Pattern
  Recognition (CVPR)}, 2017, pp. 1800--1807.

\bibitem{chang2021subspectral}
Simyung Chang, Hyoungwoo Park, Janghoon Cho, Hyunsin Park, Sungrack Yun, and
  Kyuwoong Hwang,
\newblock ``Subspectral normalization for neural audio data processing,''
\newblock {\em ICASSP 2021 - 2021 IEEE International Conference on Acoustics,
  Speech and Signal Processing (ICASSP)}, pp. 850--854, 2021.

\bibitem{Ramachandran2017SwishAS}
Prajit Ramachandran, Barret Zoph, and Quoc~V. Le,
\newblock ``Swish: a self-gated activation function,''
\newblock {\em arXiv: Neural and Evolutionary Computing}, 2017.

\bibitem{Kim2021b}
Byeonggeun Kim, Seunghan Yang, Jangho Kim, and Simyung Chang,
\newblock ``{QTI} submission to {DCASE} 2021: Residual normalization for
  device-imbalanced acoustic scene classification with efficient design,''
\newblock Tech. {R}ep., DCASE2021 Challenge, June 2021.

\bibitem{Mao2019AudioCU}
Yu~Wu, Hua Mao, and Zhang Yi,
\newblock ``Audio classification using attention-augmented convolutional neural
  network,''
\newblock {\em Knowledge-Based Systems}, vol. 161, pp. 90--100, 2018.

\bibitem{9439825}
Khaled Koutini, Hamid Eghbal-zadeh, and Gerhard Widmer,
\newblock ``Receptive field regularization techniques for audio classification
  and tagging with deep convolutional neural networks,''
\newblock {\em IEEE/ACM Transactions on Audio, Speech, and Language
  Processing}, vol. 29, pp. 1987--2000, 2021.

\end{thebibliography}

\end{document}